\documentclass{ws-ijmpb}

\def\prb{{\em Phys. Rev. }}
\def\prl{{\em Phys. Rev. Lett. }}

\begin{document}

\markboth{Luca Capriotti}
{Finite-size spin-wave theory of a collinear antiferromagnet}

%
%

\title{FINITE-SIZE SPIN-WAVE THEORY OF A COLLINEAR ANTIFERROMAGNET}
\author{LUCA CAPRIOTTI}

\address{Kavly Institute for Theoretical Physics, University of California at Santa  Barbara \\
Santa Barbara CA 93106-4030, United States of America \\
caprio@kitp.ucsb.edu}

\maketitle

\begin{history}
\received{Day Month Year}
\revised{Day Month Year}
\end{history}
\begin{abstract}
The ground-state and low-energy properties of the two-dimensional
$J_1{-}J_2$ Heisenberg model in the collinear phase are investigated using
finite-size spin-wave theory
[Q. F. Zhong and S. Sorella, {\em Europhys. Lett.} {\bf 21}, 629 (1993)],
and Lanczos exact diagonalizations.
For spin one-half -- where the effects of
quantization are the strongest --
the spin-wave expansion turns out to be
quantitatively accurate for $J_2/J_1\gtrsim 0.8$.
In this regime, both the magnetic structure factor
and the spin susceptibility are very close to the spin-wave predictions.
The spin-wave estimate of the order parameter in the collinear phase,
$m^\dagger\simeq 0.3$, is in remarkable agreement
with recent neutron scattering measurements on ${\rm Li_2VOSiO_4}$.
\end{abstract}



\section{Introduction}

The experimental realization of quasi-2d
frustrated antiferromagnets,\cite{carretta} such as ${\rm Li_2VOSiO_4}$,
${\rm Li_2VOGeO_4}$, and ${\rm VOMoO_4}$ has recently generated a renewed 
interest in the physics of the so-called {\em collinear} antiferromagnets. 
In these compounds the relevant superexchange interactions 
involve $s=1/2~~V^{4+}$ ions on weakly coupled stacked planes,
and the magnetic behavior is likely to be described by
the $J_1{-}J_2$ Heisenberg model on the square lattice,
\begin{equation} \label{j1j2ham}
{\cal{H}}=J{_1}\sum_{n.n.}
{\bf {S}}_{i} \cdot {\bf {S}}_{j}
+ J{_2}\sum_{n.n.n.}
{\bf {S}}_{i} \cdot {\bf {S}}_{j}~,
\end{equation}
where $J_1$ and $J_2$ are the (positive) nearest-neighbor ({\em n.n.}) and 
next-nearest-neighbor ({\em n.n.n.}) couplings, respectively. 
In the experimentally relevant case $J_2 \gtrsim J_1$, 
frustration is known to originate in the classical ($s\to \infty$) limit
a collinear low-temperature phase.\cite{doucot,poilblanc,ccl}
In fact, for $J_2/J_1>0.5$,  the ground state for $s\to \infty$
is a state where the spins are ferromagnetically 
aligned along one direction and antiferromagnetically in the other, 
corresponding to magnetic wave vectors $Q=(\pi,0)$ or $Q=(0,\pi)$.
These two families of states break not only the SU(2) symmetry and 
the translational invariance of the Hamiltonian, 
as the conventional N\'eel state, but also 
its invariance under $\pi/2$ real-space rotations.
Interestingly, this additional two-fold degeneracy is expected
to generate non-trivial finite-temperature properties. In fact,
an Ising-like order parameter can be defined in order
to discriminate between the two classes of states which are connected
by real space $\pi/2$ rotations, and
a finite-temperature second-order phase transition is 
expected.\cite{ccl}

How these low-temperature properties are affected by quantum
fluctuations, especially for the experimentally interesting
$s=1/2$ case, is an open problem currently preventing
a faithful comparison with the experiments.\cite{carretta,singh,misguish}
In fact, due to numerical instabilities induced by frustration
({\em sign-problem}),\cite{caprio} this issue cannot be investigated
with the powerful stochastic numerical methods successfully 
employed for the {\em n.n} Heisenberg model \cite{manousakis}, 
so that approximate
approaches have to be followed.\cite{caprio,singh,misguish}

In this paper, we address the reliability of the spin-wave ($sw$) theory
as an analytical tool to give {\em quantitative} predictions
on the low-energy properties of the $J_1{-}J_2$ Heisenberg model
in the collinear phase.
To this purpose, a previously introduced finite-size $sw$ theory \cite{zhong} 
is generalized to the collinear phase and a direct comparison
with Lanczos exact diagonalization results \cite{lanczo} is performed
for $s=1/2$.
This allows us to demonstrate the effectiveness of $sw$ theory
in describing the low-energy properties of
the $J_1{-}J_2$ Heisenberg antiferromagnet in the experimentally relevant 
regime $J_2 \gtrsim J_1$.
 
\section{Finite-size spin-wave theory}

The finite-size $sw$ theory of Zhong and Sorella,\cite{zhong}
is a rigorous generalization of the standard $sw$ approach 
to finite clusters with $N$ sites, which allows one to
avoid the spurious Goldstone-mode divergences --
related to the SU(2) symmetry breaking assumption --
in a straightforward way, without imposing any {\em ad hoc}
holonomic constraint on the sublattice magnetization.\cite{takahashi}
In the collinear case,
a systematic finite-size $sw$ expansion
can be derived for $J_2/J_1>0.5$ considering as
the approximated ground state the symmetric linear combination 
\begin{equation}\label{gs}
|\psi_{sw}\rangle = \big(|\pi,0\rangle+|0,\pi\rangle\big)/\sqrt{2},
\end{equation}
where $|\pi,0\rangle$ and $|0,\pi\rangle$ are the (normalized) $sw$ ground states
obtained assuming the classical collinear order  with 
$Q=(\pi,0)$, and $Q=(0,\pi)$, respectively.
Since on any finite size the ground state is expected to
have all the spatial symmetries of the Hamiltonian
it is important to consider the symmetric linear combination
(\ref{gs}) in order to obtain an accurate description
of its correlations. In addition, this is also crucial
to generalize rigorously the finite-size $sw$ 
theory to the collinear phase.

Both $|\pi,0\rangle$ and $|0,\pi\rangle$ in Eq.~(\ref{gs})
can be obtained according to the standard $sw$ procedure. 
The first step is to apply the unitary transformation which defines a spatially varying
reference frame pointing along the local direction of the spins in the classical state, 
assumed in the $xy$ plane:
${\cal U}=\exp[\pi/2\sum_i (\cos{Q \cdot r_i}-1) S^z_i]$. Then, using the 
Holstein-Primakoff spin-boson transformation to
first-order in $1/s$, $S^x=s-a^\dagger a$, $S^y=\sqrt{\frac{s}{2}}(a^\dagger+a)$, 
$S^z=i\sqrt{\frac{s}{2}}(a^\dagger-a)$, and translational invariance, 
the leading term of the Hamiltonian in the $sw$ expansion reads 
\begin{equation}
\label{HSW}
{\cal H}_{sw}=E_{cl} +
2J_1s\sum_{k}\Big[ A_{k} a^{\dagger}_{k}a_{k}
+ \frac{1}{2}B_{k}
(a^{\dagger}_{k}a^{\dagger}_{-k}+a_{k}a_{-k})\Big]
\end{equation}
where $E_{cl}=-2J_2Ns^2$ is the classical ground-state energy, 
and $a_k^\dagger=1/\sqrt{N}\sum_r \exp[-i k\cdot r] a_r^\dagger$, with $k=(k_x,k_y)$ 
varying in the first Brillouin zone of the lattice.
Here $A_{k}=2 J_2/J_1 +\cos{k_x}$ and $B_{k}=-(\cos{k_y}+2 J_2/J_1\cos{k_x}\cos{k_y})$
for $Q=(0,\pi)$ and with $k_x \leftrightarrow k_y$ for $Q=(\pi,0)$. 
Such leading part of the Hamiltonian, being bilinear in the Bose operators,
can be diagonalized for $k\neq (0,0), (0,\pi), (\pi,0), (\pi,\pi)$ 
using the well-known Bogoliubov transformation,
$a_{k}=u_{k}\alpha_{k}+v_{k}\alpha^{\dagger}_{-{k}}$,
with $u_{k}= \sqrt{(A_{k}+\omega_{k})/2\omega_{k}}$~, 
$v_{k}=-{\rm sgn}(B_{k}) \sqrt{(A_{k}-\omega_{k})/2\omega_{k}}$, 
$\omega_{k}=\sqrt{A^2_{k}-B^2_{k}}$ being the $sw$ dispersion relation.
In contrast, the Goldstone modes cannot be diagonalized by this transformation
since $u_k$ and $v_k$ are not defined for the $k$-vectors for which $\omega_k=0$.
For infinite systems such divergences are integrable in two dimensions and do not
lead to any contribution. On any finite-size system, instead, they are important and they 
must be treated independently. 

To this end, it is possible to define a set of Hermitian operators commuting
with each other and with the Hamiltonian: $Q_k=i(a^\dagger_k-a_k)$
for $k=(0,0)$ and $(\pi,\pi)$ and  $Q_k=(a^\dagger_k+a_k)$ 
for $k=(0,\pi)$ and $(\pi,0)$. With these definitions 
the Goldstone-mode contribution to (\ref{HSW}) reads
${\cal H}_{sm}=J_1s\sum_{k}^{sm} A_k \left[Q_k^2-1\right]$, with the sum extended only
to the four singular modes. This allows us to diagonalize
the Hamiltonian for any finite size in a basis where the $Q_k$'s have definite
quantum numbers. In particular, being $A_k>0$ for $J_2/J_1>0.5$, ${\cal H}_{sm}$
favors a ground state with $Q_k=0$ and it is easy to show that 
this implies to order $1/s$ 
$S^y(0,0)=S^z(0,0)=0$ and $S^y(\pi,\pi)=S^z(\pi,\pi)=0$ 
where 
$S^\alpha(q)= \sum_r \langle S^\alpha_0 S^\alpha_r \rangle e^{i q\cdot r}$
is the $\alpha$ component of the magnetic structure factor.
These relations together with 
the sum rules $S^x(0,0)=S^x(\pi,\pi)=0$, following 
from the translational invariance of the $sw$ Hamiltonian,
are consistent with a singlet ground-state with a zero value of the total spin 
on each sublattice. This is rigorously true according to the
Lieb-Mattis theorem \cite{lieb} for $J_2/J_1\to \infty$. In fact, in this limit
the system decouples into two independent $n.n.$ Heisenberg models 
on the two sublattices.
In general, a singlet ground state is also expected for the $J_1{-}J_2$ model
on any finite size with an even number of sites and non-frustrating boundary conditions. 
The sum rules $S^\alpha(\pi,\pi)=0$, instead, are not necessarily 
true on a finite size and for finite values of 
$J_2/J_1$ but they turn out to be fulfilled with good approximation in the
collinear phase (see below).

\begin{figure}[th]
\centerline{\psfig{figure=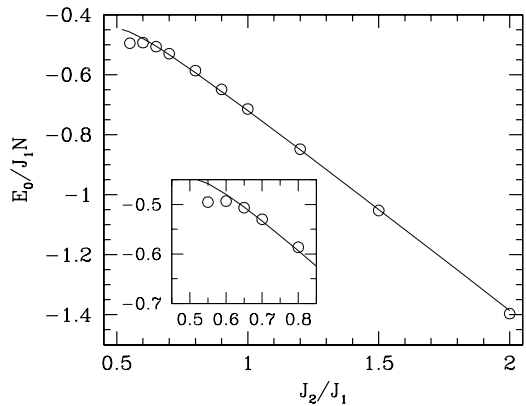,width=9cm}}
\vspace{-1.4cm}
\caption{\label{erg}
Ground-state energy as a function of the frustration ratio $J_2/J_1$,
for $N=36$.
The line is the finite-size spin-wave prediction and the empty circles are
the exact diagonalization results. The inset is an enlargement
of the strongly frustrated regime.}
\end{figure}

According to this procedure, the $sw$ estimate of the ground-state energy 
reads, $E_0=E_{cl}+J_1s\sum_k \left[\omega_k-A_k\right]$, and it turns out to be
in remarkable agreement with the exact results for $s=1/2$ and $N=36$ 
in the whole range $J_2/J_1\gtrsim0.7$ (Fig.~\ref{erg}),
i.e., far enough from the classical critical point $J_2/J_1=0.5$ separating the
collinear from the N\'eel zero-temperature phases. In fact, in this strongly frustrated regime, 
a non-magnetic ground state of purely quantum mechanical nature is likely to be stabilized
for $s=1/2$ (Ref. \refcite{note}). It is therefore natural to expect $sw$ theory
to fail in describing its correlations. Indeed, approaching the fully frustrated point $J_2/J_1=0.5$,
the $sw$ expansion eventually breaks down leading to a zero value of the antiferromagnetic 
order parameter \cite{doucot} and of the uniform susceptibility (see below).

A little more involved calculation allows us also to access the spin-rotation
invariant spin-spin correlation functions:
$C(r)=\langle {\bf S}_i\cdot {\bf S}_{i+r}\rangle$. 
The different contributions 
coming from the two degenerate states in Eq.~(\ref{gs}), $C_{Q}(r)$, 
can be found by adding to the Hamiltonian a term of the form 
${\cal H}(h_r)=J_2h_r/2 \sum_{i,\tau} {\bf S}_i\cdot {\bf S}_{i+\tau}$ 
where the vectors $\tau$ are the $z_\tau$ equivalent lattice
vectors that can be reached from a site $i$ with $|\tau|=|r|$. 
The resulting $sw$ Hamiltonian has the same form of Eq.~(\ref{HSW}) with field-dependent coefficients 
and it can be diagonalized similarly to the zero-field case.
Then, according to the Hellmann-Feynman theorem,\cite{hf} the spin-spin correlations 
can be obtained by differentiating the $sw$ 
energy in presence of the perturbation with respect to $h_r$, 
after a careful treatment of the singular modes. 
In particular,
in the collinear case, such rigorous treatment is possible
only assuming for the ground state the symmetric form (\ref{gs}).

\begin{figure}[th]
\centerline{
\psfig{figure=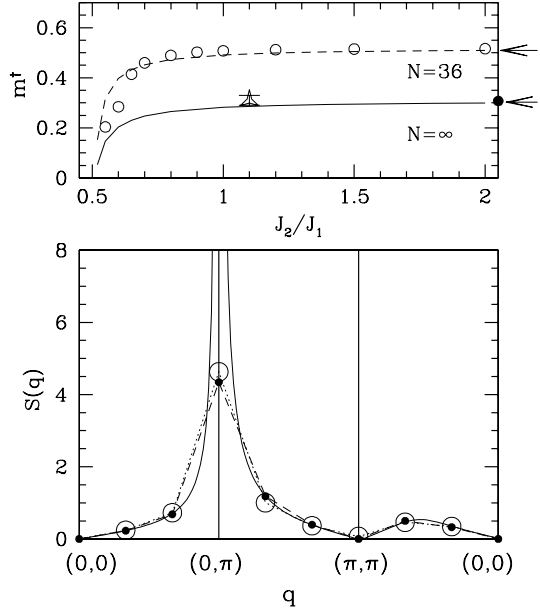,width=9cm}}
\vspace{-0.8cm}
\caption{\label{esseq}
Lower panel: magnetic structure factor for $J_2/J_1=1$, and $N=36$. Spin-wave
(full dots and dashed line), exact (empty circles and dotted line).
Upper panel: antiferromagnetic order parameter at wave vector $Q$ as a function
of the frustration ratio $J_2/J_1$, for $N=36$. Spin-wave (dashed line),
exact (empty circles). The arrows and the full dot
indicate the $sw$ and the exact \protect\cite{heis}
results for the $n.n.$ Heisenberg model, respectively, and
the star is the experimental result for
${\rm Li_2VOSiO_4}$  \protect\cite{carretta2} which has a still
undetermined $J_2/J_1 \protect\gtrsim 1$. \protect\cite{carretta,misguish,pickett}
The continuous lines in both panels are the spin-wave results in the thermodynam
ic limit.
}
\end{figure}

The resulting correlation functions have different expressions depending on the form of the vector $r$. 
In particular, for $r$ connecting the same sublattice, they read:
\begin{equation}
C_Q(r)=s^2+s\delta_{r,0} -\frac{2s}{N}\sum_k^{ns}\left(1-\cos{k\cdot r}\right)v_k^2~,
\end{equation}
if $e^{i Q\cdot r}=1$, and
\begin{equation}
C_Q(r)=-s^2-\frac{4s}{N}+\frac{2s}{N}\sum_k^{ns}\left(v_k^2-\cos{k\cdot r}\,u_kv_k\right)
\end{equation}
if $e^{i Q\cdot r}=-1$, with the sums extended to all the non singular ($ns$) 
$k$'s. 
For $r$ connecting different sublattices, instead, 
the spin-spin correlations are:
\begin{equation}
C_Q(r)=\frac{s}{N}\sum_k^{ns} (\tau_kv_k^2-\tilde{\tau}_ku_kv_k)~, 
\end{equation}
where $\tau_k=1/z_\tau\sum_\tau\cos{k\cdot \tau}$ with $e^{i Q\cdot \tau}=1$,
and $\tilde{\tau}_k=1/z_\tau\sum_\tau\cos{k\cdot \tau}$ with $e^{i Q\cdot \tau}=-1$.
The $sw$ spin-spin correlation functions calculated as
$C(r)=\left(C_{(0,\pi)}(r)+C_{(\pi,0)}(r)\right)/2$
are invariant under the symmetry operations of the lattice, as expected 
for the ground-state expectation values on any finite size.
Using these expressions the spin-isotropic magnetic structure factor, 
$S(q)= \sum_r C(r) e^{i q\cdot r}$, can be also calculated. In addition, 
having obtained an ordered expansion in $1/s$ for the spin-spin correlation functions, 
$C(r)=s^2(e^{i(\pi,0)\cdot r}+e^{i (0,\pi)\cdot r})/2 +s \alpha(r)$, 
the antiferromagnetic order parameter, $m^\dagger = \sqrt{2/NS(Q)}$,\cite{rad2}
can be expanded as $m^\dagger=s(1+\hat{\alpha}/s)$ with
$\hat{\alpha}=1/N\sum_r e^{i Q\cdot r} \alpha(r)$, and it reads:
\begin{equation}
m^\dagger = s\left[1+\frac{2}{Ns}-\frac{1}{Ns}\sum_k^{ns}v_k^2\right]~.
\end{equation}
Note that for $N\to\infty$ 
the known $sw$ result \cite{doucot,poilblanc} 
for the thermodynamic order parameter,
$m^\dagger=\sqrt{\langle 0,\pi|(S^{x}_i)^2|0,\pi\rangle }$, is recovered.
As shown in Fig.~\ref{esseq}, the $sw$ magnetic structure factor, $S(q)$, 
is in remarkable 
quantitative agreement with the exact diagonalization result for $s=1/2$ and $J_2/J_1=1$ for all 
the wave vectors $q$, and the $sw$ sum rule $S(\pi,\pi)=0$ is fulfilled
almost exactly.
In particular, the $sw$ approach provides 
a very accurate estimate (within a few percent) of the antiferromagnetic
order parameter in the whole range of $J_2/J_1 \gtrsim 0.7$.

The sharp drop of the order parameter at $J_2/J_1 \simeq 0.7$
is suggestive of a first order transition between the gapped and the
collinear phase, as also indicated by a recent variational 
study.\cite{var} This leads in particular to a very fast 
saturation of the order parameter to the value of the $n.n.$ 
Heisenberg model,\cite{heis} expected for $J_2/J_1\to\infty$.
Remarkably, the $sw$ prediction for
$J_2/J_1 \gtrsim 1.0$ is in excellent agreement
with the result of recent neutron scattering  
measurements \cite{carretta2} on ${\rm Li_2VOSiO_4}$, 
giving $m^\dagger=0.31(2)$ in the zero temperature limit. This compound is known
to have $J_2\gtrsim J_1$, even if the precise value 
of the frustration ratio is presently much debated.
\cite{carretta,misguish,pickett}
Unfortunately, the weak dependence of the order parameter 
on $J_2/J_1$ for $J_2/J_1 \gtrsim 1.0$
does not allow to use the $sw$ prediction 
to determine more precisely this ratio.

Within the finite-size $sw$ theory, it is also possible to study
the structure of the low-energy spin excitation spectrum which is connected
to the ground-state antiferromagnetic correlations and to the uniform
susceptibility.
Following Lavalle {\em et al.},\cite{lavalle} in order to 
stabilize an excitation of total spin $S$, a Zeeman term,
 ${\cal H}(h)=-hs\sum_{i}{S}_{i}^{z}$, is added
to the spin Hamiltonian.
Classically, for magnetic fields not large enough to induce a spin-flop transition,
the classical solution is simply obtained by canting the spins of 
an angle $\theta_h$ along the direction of the field $h$ [$\sin\theta_h=h/4J_1(1+2J_2/J_1)$].
After a new rotation of the reference frame in order to align it to the new directions of the spins,
the finite-size $sw$ expansion is straightforward and again it leads to
a linearized Hamiltonian of the same form of Eq.~(\ref{HSW})
with field-dependent coefficients 
$A^h_{k}=A_k+B_k\sin^2\theta_h$, $B^h_{k}=B_k(1-\sin^2\theta_h)$, and classical energy.
In this case, due to the reduced spin-rotation symmetry of the Hamiltonian 
only the $(0,0)$ and the $(\pi,\pi)$ modes are singular. 
The former favors a value of $S^z(0,0)=N (s \sin\theta)^2$ consistent with the applied field,
at the classical level, while the latter generates 
the usual approximated sum rule $S^z(\pi,\pi)=0$, which is not affected 
by the uniform field along the $z$ direction.
The total spin $S=N \langle S^z_{i} \rangle$ of the
excitation can be related to the magnetic field $h$ by means of
the Hellmann-Feynman theorem
\begin{eqnarray}
\langle S^z_{i} \rangle_h&=&-\frac{1}{Ns}\frac{\partial}{\partial h}E(h) 
=s\frac{h}{4J_1(1+2J_2/J_1)} \times \nonumber \\
&& \left[1-\frac{1}{2Ns(1+2J_2/J_1)} \sum_{k}^{ns} B_k
\sqrt{\frac{A^h_{k}+B^h_{k}} {A^h_{k}-B^h_{k}}}\right]~,
\end{eqnarray}
where 
\begin{equation} \label{eh}
E(h)=E_{cl}-(sh)^2\frac{N}{8J_1(1+2J_2/J_1)}+J_1s\sum_{k} \left[\omega_k^h-A_k^h\right]~,
\end{equation}
and $\omega_{k}^h=\sqrt{(A^h_{k})^2-(B^h_{k})^2}$.
Using these expressions the energy spectrum $E(S)$ can be calculated by means of 
a Legendre transform, $E(S)=E(h)+hsS$. Besides, the 
uniform susceptibility in the thermodynamic limit,
$\chi=-1/Ns^2 \,{\partial^2 E(h)}/{\partial h^2}|_{h=0}$,
can be obtained from Eq.~(\ref{eh}) by direct derivation and it reads:
\begin{equation}
\label{eq.chisw}
\chi/\chi_{cl}=1-\frac{1}{2s(1+2J_2/J_1)}\int \frac{{\rm d}^2k}{(2\pi)^2}\,B_k
\sqrt{\frac{A_{k}+B_{k}}{A_{k}-B_{k}}}~.
\end{equation}
where $\chi_{cl}=1/4J_1(1+2J_2/J_1)$ is the susceptibility in 
the classical limit.

\begin{figure}[th]
\centerline{
\psfig{figure=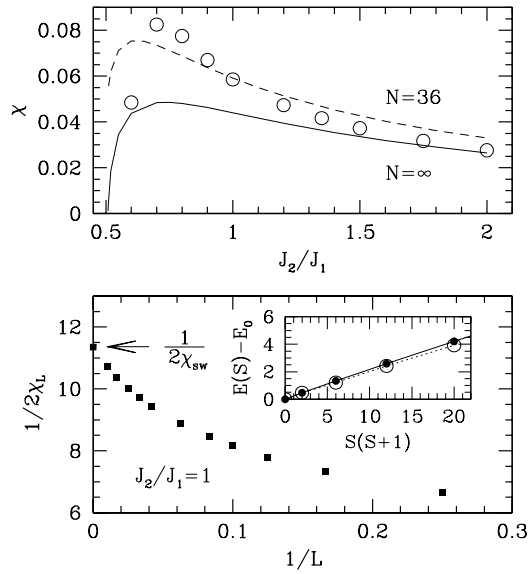,width=9cm}}
\vspace{-1.0cm}
\caption{\label{chisw}
Lower panel: size dependence of the spin-wave $1/2\chi_L$ (see text)
for $J_2/J_1=1$. Inset: spin-wave
(full dots and continuous line) and exact (empty circles and dotted line)
low-energy spectra as a function
of $|{\bf S}^2|=S(S+1)$ for $N=36$, and $J_2/J_1=1$.
Upper panel: uniform susceptibilities as a function of $J_2/J_1$.
Dashed line: spin-wave $\chi_{S=1}$ for $N=36$. Empty circles: exact $\chi_{S=1}
$ for $N=36$.
Continuous line: thermodynamic spin-wave $\chi$.
}
\end{figure}

Due to the discreteness of the energy spectrum, a direct comparison of the
$sw$ susceptibility with the exact diagonalization results is not possible
on a finite size. However, in a quantum antiferromagnet with a long-range 
ordered ground state, the susceptibility 
is directly related to the properties of the low-energy spectrum 
and this can be exploited to establish the accuracy of the $sw$ 
predictions. 
In fact, whenever long-range order is present in the 
thermodynamic limit, the low-lying spin excitations ($S \ll \sqrt{N}$)
order themselves as in the spectrum of a free quantum rotator,  
$E(S)-E_0=S(S+1)/2IN$, where $I$ is 
the so-called {\em momentum of inertia} per site. 
This definition is very close to the definition of the uniform spin
susceptibility, which
can be calculated by taking first the infinite-volume limit of the energy 
per site at fixed magnetization $m=S/N$, $e(m)=E(S)/N$, and then letting 
$m\to 0$: $e(m)=e_0+m^2/(2\chi)$. As it is known from the low-energy theory 
of a quantum antiferromagnet,\cite{nlsm} this allows for an identification between $I$ and $\chi/N$
so that the quantity 
$\left[ 2 \chi_{S} \right]^{-1} = N\left(E(S)-E_0\right)\left[S(S+1)\right]^{-1}$
approaches for infinite size the physical inverse susceptibility $1/2\chi$, for any spin-excitation $S\ll N$. 
This features are clearly verified by the $sw$ scheme
as it is illustrated in the lower panel of Fig.~\ref{chisw} for $J_2/J_1=1$,
where $1/2\chi_S$ is plotted for $S=L \equiv \sqrt{N}$ and approaches
the thermodynamic value calculated according to
Eq.~(\ref{eq.chisw}) ($1/2\chi_{sw}\simeq11.36$).

The comparison with the exact results of the $sw$ 
spectrum $E(S)-E_0$ vs $S(S+1)$ (inset of Fig.~\ref{chisw}), and, equivalently,
of the quantity $1/2\chi_{S=1}$ (upper panel of Fig.~\ref{chisw}) 
demonstrates the accuracy of $sw$ theory in estimating the spin susceptibility
of the $J_1{-}J_2$ model in the collinear phase for $J_2/J_1\gtrsim 0.8$. 
As it is shown in the same figure (upper panel),
the classical uniform susceptibility is strongly renormalized
by quantum fluctuations for $s=1/2$ at the $sw$ level.
As expected, approaching $J_2/J_1=0.5$ such reduction is enhanced
due to the increased frustration and it leads eventually to the 
vanishing of the susceptibility for $J_2/J_1\simeq 0.507$ and to the expected transition 
to a spin-gapped non-magnetic phase.\cite{caprio} However, from the comparison with the
exact diagonalization results for both the uniform susceptibility and the order parameter, 
it appears clear that the effects of quantum fluctuations are 
underestimated by the $sw$ expansion in the regime of strong frustration so that the
transition to the non-magnetic region is likely to occur at 
slightly higher values of the $J_2/J_1$ ratio ($J_2/J_1\simeq 0.6$).

\section{Conclusions}

In conclusion, we have presented a finite-size spin-wave
study of the $J_1{-}J_2$ Heisenberg antiferromagnet in the collinear phase,
focusing in particular on the ground-state spin-spin correlations
and the low-energy excitations.
For $s=1/2$, the comparison with 
exact diagonalization results reveals a remarkable agreement
for $J_2/J_1\gtrsim 0.8$.
In particular, both the antiferromagnetic structure factor
and the spin susceptibility are very close to the spin-wave predictions.
The accuracy of the results on finite sizes
indicates that the spin-wave expansion provides 
a quantitatively reliable description of the ground state
of the $J_1{-}J_2$ model in the experimentally relevant
regime $J_2\gtrsim J_1$. This is confirmed by
recent neutron scattering measurements on ${\rm Li_2VOSiO_4}$,
providing a value of the order parameter in excellent agreement 
with the spin-wave prediction in the
collinear phase, $m^\dagger\simeq 0.3$. 

The accuracy of spin-wave theory in describing
the low-energy excitations also suggests that theoretical approaches
to the thermodynamics based on a Gaussian treatment of
quantum fluctuations, such as effective Hamiltonian methods, 
\cite{pqscha} are likely to successfully describe 
the finite-temperature properties of the $J_1{-}J_2$ model
in the collinear phase. Work on this line
of research would allow to establish a closer contact between the
frustrated Heisenberg model and the physical properties of the 
recently synthesized collinear antiferromagnets, helping in
their precise characterization and in the determination of
the actual frustration ratios.

I wish to thank Alberto Parola, Sandro Sorella, and Valerio Tognetti 
for their teachings, support, and continuous encouragement. 
Special thanks to Pietro Carretta for a very useful correspondence
and for kindly communicating the experimental results on ${\rm Li_2VOSiO_4}$
before publication. Thanks to Federico Becca 
for help in performing the Lanczos calculations, and 
to Frederic Mila, and Doug Scalapino for many fruitful discussions.
Special thanks also to Liz Osterberg for a careful reading of the manuscript.
CPU time kindly provided by SISSA and INFM-Democritos is 
gratefully acknowledged.
This work was supported by NSF under grant DMR-9817242.



\begin{thebibliography}{0}


\bibitem{carretta} R. Melzi, P. Carretta, A. Lascialfari, M. Mambrini, M. Troyer,
P. Millet, and F. Mila, \prl {\bf 85}, 1318 (2000); 
R. Melzi, S. Aldovrandi, F. Tedoldi,
P. Carretta, P. Millet, and F. Mila, \prb {\bf B64}, 024409 (2001).
P. Carretta, R. Melzi,
N. Papinutto, and P. Millet, \prl {\bf 88}, 47601 (2002);
P. Carretta, N. Papinutto, C. B. Azzoni, M.C. Mozzati, E. Pavarini, S. Gonthier, and 
P. Millet, cond-mat/0205092.

\bibitem{doucot} P. Chandra and B. Doucot, \prb {\bf B38}, 9335 (1988).

\bibitem{poilblanc} H. J. Schulz, T. A. L. Ziman, and D. Poilblanc,
{\em J. Phys. I (France)} {\bf 6}, 675 (1996).

\bibitem{ccl} P. Chandra, P. Coleman, and A. I. Larkin, \prl {\bf 64}, 88 (1990).

\bibitem{singh} R. R. P. Singh, W. Zheng, J. Oitmaa, O. P. Sushkov, and C. J. Hamer, cond-mat/0303075.

\bibitem{misguish} G. Misguich, B. Bernu, and L. Pierre, cond-mat/0302583.

\bibitem{caprio} See for instance: L. Capriotti, {\em Int. J. Mod. Phys.} {\bf B15}, 1799 (2001).

\bibitem{manousakis} E. Manousakis, {\em Rev. Mod. Phys.} {\bf 63}, 1 (1991).

\bibitem{zhong} Q. F. Zhong and S. Sorella, {\em Europhys. Lett.}  {\bf 21}, 629 (1993).

\bibitem{lanczo} C. Lanczos, J. Res. Nat. Bur. Stand. {\bf 45}, 255.
See also: W. D. Joubert in {\em Siam J. on Matrix Anal. and Appl.}, Volume 13, 
Number 3, July 1992.

\bibitem{takahashi} M. Takahashi, \prb {\bf B40}, 2494 (1989); 
J. E. Hirsch and S. Tang, \prb {\bf B40}, 4769 (1989); H. Nishimori and
Y. Saika, {\em J. Phis. Soc. Jpn.} {\bf 59}, 4454 (1990).  


\bibitem{lieb} E. Lieb and D. Mattis, {\em J. Math. Phys.} {\bf 3}, 749 (1962).

\bibitem{note} The nature of the non-magnetic phase is presently
a debated issue. See for instance:
L. Capriotti, F. Becca, A. Parola, and S. Sorella,
\prl {\bf 87}, 097201 (2001); \prb {\bf B67}, 212402 (2003); 
V.N. Kotov , J. Oitmaa, O.P. Sushkov, and Z. Weihong, {\em Phil. Mag.  B}, 
{\bf 80} 1483 (2000). 

\bibitem{hf} R.P. Feynman, \prb {\bf 56}, 340 (1939).

\bibitem{rad2} The factor 2 appearing in this definition is 
related to the two-fold degeneracy of the $(0,\pi)$ and $(\pi,0)$
states,\cite{poilblanc} and allows one to 
recover the $n.n.$ Heisenberg limit for $J_2/J_1\to \infty$.

\bibitem{var} A. Parola,  F. Becca, L. Capriotti, and S. Sorella,
{\em J. Mag. Mag. Mat.}, to be published.

\bibitem{heis}  A.W. Sandvik, \prb {\bf B56}, 11678 (1997);
M. Calandra Buonaura and S. Sorella, \prb {\bf B57}, 11446 (1998).

\bibitem{carretta2} A. Bombardi, L. Paolasini, P. Carretta,
J. Rodriguez-Carvajal, P. Millet, and R. Caciuffo, communicated by
P. Carretta.

\bibitem{pickett} H. Rosner, R. R. P. Singh, W. H. Zheng, J. Oitmaa,
S.-L. Drechsler, and W. E.  Pickett, \prl {\bf 88}, 186405 (2002).  

\bibitem{lavalle} C. Lavalle, S. Sorella and A. Parola, \prl {\bf 80}, 1746 (1998).

\bibitem{nlsm} S. Chakravarty, B.I. Halperin, and D.R. Nelson, \prl {\bf 60}, 1057 (1988)


\bibitem{pqscha} A. Cuccoli, V. Tognetti, R. Vaia, and P. Verrucchi,
\prl {\bf 77}, 3439 (1996).


\end{thebibliography}
\end{document}